\documentclass[letterpaper]{article} 
\usepackage{aaai21}  
\usepackage{times}  
\usepackage{helvet} 
\usepackage{courier}  
\usepackage[hyphens]{url}  
\usepackage{graphicx} 
\usepackage{natbib}  
\usepackage{caption} 
\usepackage{multirow}
\usepackage{multicol}
\usepackage{hhline}
\usepackage{caption}
\usepackage{subcaption}
\usepackage{pifont}
\usepackage{makecell}
\usepackage{tabu}
\usepackage{amssymb}
\usepackage{amsmath}

\RequirePackage[prologue]{xcolor}
\definecolor[named]{ACMBlue}{cmyk}{1,0.1,0,0.1}
\definecolor[named]{ACMYellow}{cmyk}{0,0.16,1,0}
\definecolor[named]{ACMOrange}{cmyk}{0,0.42,1,0.01}
\definecolor[named]{ACMRed}{cmyk}{0,0.90,0.86,0}
\definecolor[named]{ACMLightBlue}{cmyk}{0.49,0.01,0,0}
\definecolor[named]{ACMGreen}{cmyk}{0.20,0,1,0.19}
\definecolor[named]{ACMPurple}{cmyk}{0.55,1,0,0.15}
\definecolor[named]{ACMDarkBlue}{cmyk}{1,0.58,0,0.21}

\newcommand{\centered}[1]{ \begin{tabular}{l} \\  #1 \\\vspace{0.5pt}\end{tabular}}
\newcommand{\red}[1]{\textcolor{ACMRed}{#1}}
\newcommand{\yellow}[1]{\textcolor{ACMYellow}{#1}}
\newcommand{\green}[1]{\textcolor{ACMGreen}{#1}}
\newcommand{\no}{\red{\ding{55}}}
\newcommand{\yes}{\green{\bf \checkmark}}
\newcommand{\low}{\red{Low}}
\newcommand{\medium}{\yellow{Medium}}
\newcommand{\high}{\green{High}}

\title{Measuring Disparate Outcomes of Content Recommendation Algorithms with Distributional Inequality Metrics}

\author {
    Tomo Lazovich\textsuperscript{\rm 1}\footnote{Email: tlazovich@twitter.com},
    Luca Belli\textsuperscript{\rm 1},
    Aaron Gonzales \textsuperscript{\rm 1},
    Amanda Bower
    \textsuperscript{\rm 1},
    Uthaipon Tantipongpipat
    \textsuperscript{\rm 1},
    Kristian Lum
    \textsuperscript{\rm 1},
    Ferenc Huszar
    \textsuperscript{\rm 2}\footnote{Work completed while author was a consultant for Twitter},
    Rumman Chowdhury
    \textsuperscript{\rm 1}\\
}

\affiliations {
    \textsuperscript{\rm 1} 
    Twitter, Inc. \\
    \textsuperscript{\rm 2} University of Cambridge\\
}

\begin{document}
%

\maketitle
\begin{abstract}
\begin{quote}
The harmful impacts of algorithmic decision systems have recently come into focus, with many examples of systems such as machine learning (ML) models amplifying existing societal biases. Most metrics attempting to quantify disparities resulting from ML algorithms focus on differences between groups, dividing users based on demographic identities and comparing model performance or overall outcomes between these groups. However, in industry settings, such information is often not available, and inferring these characteristics carries its own risks and biases. Moreover, typical metrics that focus on a single classifier's output ignore the complex network of systems that produce outcomes in real-world settings. In this paper, we evaluate a set of metrics originating from economics, \textit{distributional inequality metrics}, and their ability to measure disparities in content exposure in a production recommendation system, the Twitter algorithmic timeline. We define desirable criteria for metrics to be used in an operational setting, specifically by ML practitioners. We characterize different types of engagement with content on Twitter using these metrics, and use these results to evaluate the metrics with respect to the desired criteria. We show that we can use these metrics to identify content suggestion algorithms that contribute more strongly to skewed outcomes between users. Overall, we conclude that these metrics can be useful tools for understanding disparate outcomes in online social networks.
\end{quote}
\end{abstract}

\section{Introduction}

Content recommendation algorithms live at the heart of social media platforms, with machine learning models recommending and ranking
everything from accounts to follow, topics of interest, and the actual posts that appear in a user's feed. Over the past decade, it has become clear that attention and engagement on social media platforms is highly concentrated, with a small set of users receiving the lion's share of attention~\cite{mccurley2008income,zhu2016attention}. Understanding this skew is crucial not only for those seeking to monetize content on the internet, but, more importantly, for historically oppressed voices and social movements who have used social media to organize and find their communities. Additionally, users' trust can erode over time if they feel they are posting ``into the void," with no one seeing their content~\cite{mcclain_widjaya_rivero_smith_2021}. 

At the same time, there has been a newly increased focus on the potential harms caused by machine learning systems~\cite{noble2018algorithms,benjamin2019race,buolamwini2018gender}. A large suite of fairness metrics has emerged, mainly focused on comparing disparities in classification model performance across demographic groups~\cite{mitchell2021algorithmic,barocas2017fairness,mehrabi2021survey}. While significant advancements have been made, these group-comparison metrics still suffer from many drawbacks. Previous work has noted that these metrics struggle to be fair for intersectional groups, and that algorithms can satisfy group fairness while still being quite unfair to individuals~\cite{kearns2018preventing,dwork2012fairness}. Additionally, there are many hurdles to operationalizaton of these metrics. First, they require knowledge of a sensitive attribute in order to identify potentially disadvantaged groups.  Such demographic information is often noisy or deliberately not collected due to privacy or legal concerns. Second, they are largely targeted at analyzing single classification models, comparing performance measures like false positive rate and accuracy across the groups. Typical deployed models are performing more complex tasks, like ranking or recommendation, making these classification measures ineffective. Finally, end-user outcomes in production systems are usually the product of a large, interconnected system of models that feed into one another. In industry settings, these issues make metrics like demographic parity difficult to implement in practice. As such, there is a strong need for metrics that capture disparities in outcomes for users while simultaneously overcoming the limitations of single-model, group comparison metrics. 

\subsection{Summary of contribution}

In this work, we study the usability of a suite of metrics originally used by economists to quantify income inequality, which we refer to as \textit{distributional inequality metrics}~\cite{trapeznikova2019measuring}. These inequality metrics help overcome the previously described limitations of group fairness metrics since they do not rely on knowledge of demographic information and  focus on distribution of outcomes across an entire population. We evaluate, both qualitatively and quantitatively, the application of these metrics in the context of unequal outcomes on social media. Specifically, we 

\begin{itemize}
\item Define a set of desirable criteria for metrics attempting to capture skews in outcomes in real-world settings
\item Quantify skews in engagements with posts on the Twitter algorithmic timeline, specifically focused on disparities between the authors
\item Identify algorithmic sources of distributional inequality using the proposed metrics, observing a connection between number of followers and exposure which indicates that out of network content suggestions are more skewed for users with fewer followers
\item Use the empirical results from applying the metrics in the cases above to evaluate them in the context of the desirable criteria we set forth
\end{itemize} 

\section{Prior work}

Many researchers have investigated applying principles from economics to measure algorithmic fairness. In one recent study, a set of generalized entropy measures and inequality indices were used to quantify differences in group and individual fairness~\cite{speicher2018unified}. Another work created puppet Twitter accounts and used the Gini index to compare the diversity of authors seen by accounts using the Twitter algorithmic timeline versus those that were using a simple chronological feed of content~\cite{bandy2021more}. Researchers at LinkedIn showed that the Atkinson index could be used to promote more equitable design choices when used in A/B testing ~\cite{saint2020fairness}. More generally, there has been much work by economists and social scientists to frame the questions surrounding algorithmic fairness~\cite{rambachan2020economic,cowgill2020algorithmic}. One particularly relevant work emphasizes that equality and power are often better ways to frame questions of algorithmic harms, and it provides a basis for new metrics for computing quantities relevant to these issues from economic theory~\cite{kasy2021fairness}.

Outside of the field of economics, there has also been a focus on the fairness of ranking and recommendation, which often require different metrics compared to classification models. Several studies frame the fairness of rankings in terms of the allocation of exposure~\cite{singh2018fairness,Biega_2018,Zehlike_2017}, highlighting the need for metrics that capture disparities in attention garnered by creators in ranking contexts. It has also been shown that the notion of individual fairness~\cite{dwork2012fairness} can be applied to ranking models, extending the previous work on exposure to enforce that similar items from minority and majority groups appear together ~\cite{bower2021individually}. In these cases, the works studied the more complex ranking task, but were still focused on single models.

Our work is unique compared to previous work in multiple respects. First, we undertake a systematic evaluation of a number of different inequality metrics, many of which have not been studied previously in the context of recommendation systems or social media. Second, we apply the measurement of these metrics to flag specific algorithmic sources of skew in end-user outcomes, directly demonstrating how they can be used in a real-world operational setting on production-scale datasets.

\section{Moving from measuring model-level fairness to system-level outcomes}

While current group-based fairness metrics are effective at identifying imbalances between demographics in model performance, they do not capture how these imbalances ultimately cascade and disproportionately harm or benefit end users. Focusing on end outcomes allows us to recenter our measurements on notions of social hierarchy, power distribution, and equality~\cite{kasy2021fairness,green2021impossibility}. Researchers at LinkedIn have previously advocated for a similar approach by incorporating the Atkinson index as a measure of inequality of outcomes during A/B testing, and in this work we hope to generalize that approach to a broader set of social media use cases~\cite{saint2020fairness}. Diagnosing these issues is the first step towards what has been proposed as a more substantive approach to fairness, with the end goal being ``reforms [that] can address the forms and mechanisms of inequality that were identified" and a reckoning with the algorithmic role in these mechanisms~\cite{green2021impossibility}. 

We focus here on disparities in the distribution of exposure on Twitter as a concrete, measurable system-level outcome. While engagement is most definitely not the only measure of value gained from the platform~\cite{milli2021optimizing}, those sharing content on platforms with recommendation systems depend crucially on how they are ranked. Within social media, mobilizing social movements, showcasing work while searching for jobs, or building an audience for one's content are all examples of use cases heavily driven by level of exposure. On other platforms, level of exposure can be a direct tie to revenue, such as for sellers on eBay and Amazon, artists on Spotify, or actors and producers on Netflix. Therefore, we see the number of engagements, a measure of how much a user's content is exposed to others, as a good starting point for assessing the usefulness of these metrics. 

In this frame of mind, it is important to acknowledge that many of the mechanisms of inequality present in algorithm-driven systems will not be identifiable by metrics alone. Measuring the value of a metric inherently assumes that the disparity or harm in question is quantitatively measurable. Many classes of harms will not be capturable with a metric. This is an additional motivation for exploring different types of engagement as a testing ground, since those outcomes are quantifiable and therefore disparities should be measurable by an actionable metric. At the same time, we note that any absence of disparities measured by this metric does not imply that a system is perfectly fair, but rather that there was negligible measurable inequality for the specific quantity being studied. 

\subsection{What makes a good system-level metric?}

In order to evaluate them thoroughly, we must first define desirable criteria for metrics attempting to capture gaps in user experiences. We classify these desirable criteria in three categories: theoretical, qualitative, and empirical. Theoretical criteria are inherent mathematical properties that may be useful when deploying the metric. Qualitative criteria are criteria that are subjective evaluations of the metric. Empirical criteria are criteria for which we can use data to measure their efficacy. We evaluate metrics with respect to the following attributes, all of which contribute to their implementability in an operational context.

\subsubsection{Theoretical criteria}

Some desirable mathematical properties for criteria have been enumerated in previous economics literature~\cite{allison1978measures}, as well as LinkedIn's study of the Atkinson index~\cite{saint2020fairness}. These include:

\begin{itemize}
    \item \textbf{Population invariance}: because numbers of users frequently fluctuate, and we may want  to compare between subgroups on the platform, the metric should have the same meaning for populations of different sizes.
    \item \textbf{Adjustability}: the metric should allow practitioners to focus on different percentiles of the distribution, as a non-adjustable metric may not weight segments of the distribution appropriately for the application area being studied. 
    \item \textbf{Scale invariance}: if every value in the population is multiplied by a constant factor, the value of the metric should remain equal to its previous value
    \item \textbf{Subgroup decomposability}: the metric should be easily calculable in terms of values of the metric computed on subgroups of the population
\end{itemize}

\subsubsection{Qualitative criteria}

\begin{itemize}
    \item \textbf{Interpretability}: changes in the metric should be understandable by non-experts
    \item \textbf{User focus}: the metric should be directly tied to real people's experiences and measured in units that are easily translated into a description of a property of that population.
\end{itemize}

\subsubsection{Empirical criteria}

\begin{itemize}
    \item \textbf{Stability}: if resampling with a different population, the metric should have a similar value for similar distributions.
    \item \textbf{Effect size}: when there are changes in the skew of the distribution, differences in the metric should be large enough to be distinguishable. Said another way, distributions that are very different should show a large dynamic range in values of the metric. 
\end{itemize}

\subsection{Distributional inequality metrics: definitions}

\begin{table*}[h]
    \centering
    {\tabulinesep=1mm
    \begin{tabular}{|c|c|c|c|c|c|}
        \hline
         \bf Category & \bf Metric & \bf Definition & \bf Range & \bf Notes  \\ \hline
         \multirow{2}{*}[-3pt]{\bf Entropy} & Gini index & 
         \centered{$\frac{\sum_{p=1}^{K}\sum_{q=1}^{K}\left|V_p - V_q\right|}{2K\sum_{i=1}^{K} V_i}$} &
         $[0, 1]$ & \makecell{\small Value of 0 means perfect equality}\\ \cline{2-5}
         
         & Atkinson index & $1 - \frac{1}{\mu}\left(\frac{1}{K}\sum_{i=1}^{K} V_i^{1-\epsilon}\right)^{1/(1-\epsilon)}$ & $[0, 1]$ & \makecell{\small $\epsilon \in [0, \infty)$ is "aversion to inequality"\\ \small For $\epsilon=1$, the term divided by $\mu$ is defined \\ \small as the geometric mean of the values $V_i$. \\ \small Value of 0 means perfect equality. \\\small Goes to $1$ as $\epsilon \to \infty$. \\ \small $\mu$ is the arithmetic mean of the all values $V_i$} \\ \hhline{|=|=|=|=|=|}
         
         \multirow{2}{*}[-3pt]{\bf Ratio} & Percentile ratio & \centered{$\frac{V_{a} \text{ s.t. } P(V_{a} \leq p_a)}{V_{b} \text{ s.t. } P(V_{b} \leq p_b)}$} & $[0, \infty)$ & 
         \multirow{2}{*}{\makecell{\small $p_a$ and $p_b$ are percentiles to compare }} \\ \cline{2-4}
         
         & Share ratio & \centered{$\frac{\sum_{i=f_i(p_b)}^{K} V_i}{\sum_{j=0}^{f_i(p_a)} V_j}$} & $[0, \infty)$ & \\ \hhline{|=|=|=|=|=|}
         
         \textbf {Tail share} & Share of top X\% & \centered{$100 \times \frac{\sum_{i=f_i(1-p)}^{K} V_i}{\sum_{j=0}^K V_j}$} & $[0, 100]$ & \makecell{\small $p = X/100$} \\ \hhline{|=|=|=|=|=|}
         
         \multirow{2}{*}[-2pt]{\bf Equivalence}& \% of equal share & \centered{$p \text{ s.t. } \frac{\sum_{i=f_i(p)}^{K} V_i}{\sum_{j=0}^{f_i(p)} V_j} = 1$} & $(0, 100)$ & \\ \cline{2-5}
         
         & Equivalent to top $X$\% & \centered{\makecell{$q \text{ s.t. }$ \\  $\sum_{j=0}^{f_i(q)} V_j = \sum_{i=f_i(1-p)}^{K} V_i$}} & $(0, 100)$ & \makecell{\small $p = X/100$} \\ \hline
    \end{tabular}
    }
    \vspace{2pt}
    \caption{Mathematical definitions of distributional inequality metrics. $V_i$ is the value for population member $i$, and $K$ the total number of members of the population. For notational purposes, we assume that the values ${V_i}$ are ordered in ascending order ($V_i \leq V_j$,  $\forall (i,\ j) \text{ s.t. } i < j$). $f_i(p)$ is the index of the value at the $p$-th percentile.}
    \label{tab:metrics}
\end{table*}

In this work, we have chosen to focus on metrics that were originally used to measure the concentration of income distribution in economics, since this family of metrics is already concerned with measuring disparities across a population. These metrics capture the skew or top-heaviness of a distribution and can be applied to any set of non-negative input values. Table~\ref{tab:metrics} shows the mathematical definitions of each metric we consider. The metrics we evaluate in this study include:

\begin{itemize}
    \item \textbf{Gini index}: the Gini index can be defined as the mean absolute difference between all distinct pairs of people in the population divided by the mean value over the population~\cite{gini1912variabilita,farris2010gini}. In essence, it is a measure of how pairwise differences in income compare to the average income. It ranges from zero to one, with zero indicating perfect equality and one indicating a totally concentrated distribution.

    \item \textbf{Atkinson index}: the Atkinson index was introduced to address some limitations of the Gini index, namely that many believed it did not adequately weight those in the low part of the population ~\cite{atkinson1970measurement}. Atkinson introduces a new ``aversion to inequality" parameter, $\epsilon$, that explicitly allows for adjusting the weight of the low end of the distribution. It also ranges from 0 to 1, with higher values indicating more skew. As $\epsilon \to \infty$, the value of Atkinson index approaches 1. This is because as we weigh the low end of the distribution more, any non-equal distribution will be considered more and more disparate. 
    
    \item \textbf{Percentile ratio}: a percentile is the value at which some percentage of people have incomes less than or equal to that value. For example, if the value of the 20th percentile of a distribution \$100, then 20\% of people make less than or equal to \$100. The percentile ratio is defined as the ratio of two different percentile values. In the economics literature, this is typically used to compare the low and high ends of the distribution, most usually the ratio of the 90th and 10th or the 80th and 20th percentiles~\cite{trapeznikova2019measuring}. 
    
    \item \textbf{Share ratio}: while percentile ratios are comparing single values at particular positions in the distribution, the share ratio compares cumulative portions of the distribution. For example, the 80/20 share ratio compares the share of wealth held by people in the top 20\% (80th percentile and above) of the distribution to those in the bottom 20\%. 
    
    \item \textbf{Percentage share of top or bottom $X$\%}: while ratios are useful in capturing the scale of disparities, they don't offer any information about the values that went into the ratio. Sometimes it can be useful to directly report the share of wealth held by the top or bottom of a population. One work has suggested that while Gini alone is insufficient in capturing differences in countries' income distributions, a combination of top 10\% share, bottom 10\% share, and Gini can provide more information about disparities~\cite{sitthiyot2020simple}. 
    
    \item \textbf{Percentage of equal share}: this percentage is the percentage of people for which the bottom end of the distribution has an equal share (50\%) of the wealth as the top end of the distribution. For example, if the percentage of equal share were 99\%, that would mean that the bottom 99\% of people had a share of wealth equal to the top 1\%. 
    
    \item \textbf{Equivalent to top $X$\%}: one way to compare ends of the distribution without requiring ratios is to find equivalences in the distribution. For example, if the top 1\% of a population has 30\% of the wealth, we can ask what percentage of people at the bottom of the distribution carries the same fraction of wealth. The larger that percentage is, the larger the disparity between the high and low ends of the distribution.

\end{itemize}

Many of these metrics are also related to a distribution visualization known as the Lorenz curve, a measure of cumulative fraction of wealth as a function of cumulative population size~\cite{lorenz1905methods}. When curves don't intersect, a curve with more area under it indicates a more equitable distribution, and a diagonal line corresponds to the case where all members of the population receive the same income. See the appendix for details on the visual interpretation of the Lorenz curve, as well as its relationship to the metrics defined above.

For organizational purposes, we can sort the metrics into four different categories. The Atkinson index is part of a family of more generalizable entropy measures~\cite{shorrocks1980class}, and Gini also measures an entropy-like quantity~\cite{gini_entropy}, so we will refer to them together as `entropy' metrics. We define the second group as `ratio' metrics, containing the percentile and share ratios. The third group is defined as `tail share' metrics, and it contains the more general non-ratio measures of share of wealth at the top or bottom of the distribution. The final class of metrics is `equivalence' metrics, or those metrics defining percentiles where shares of wealth distribution are equal to one another. Now, having defined these metrics, we measure them on different types of engagements with content on Twitter. We defer discussion of the metrics with respect to the criteria set out above until later sections, in order to be able to evaluate theoretical properties together with empirical and qualitative ones.

\begin{figure*}
     \centering
     \begin{subfigure}[b]{0.4\textwidth}
         \centering
         \includegraphics[width=\textwidth]{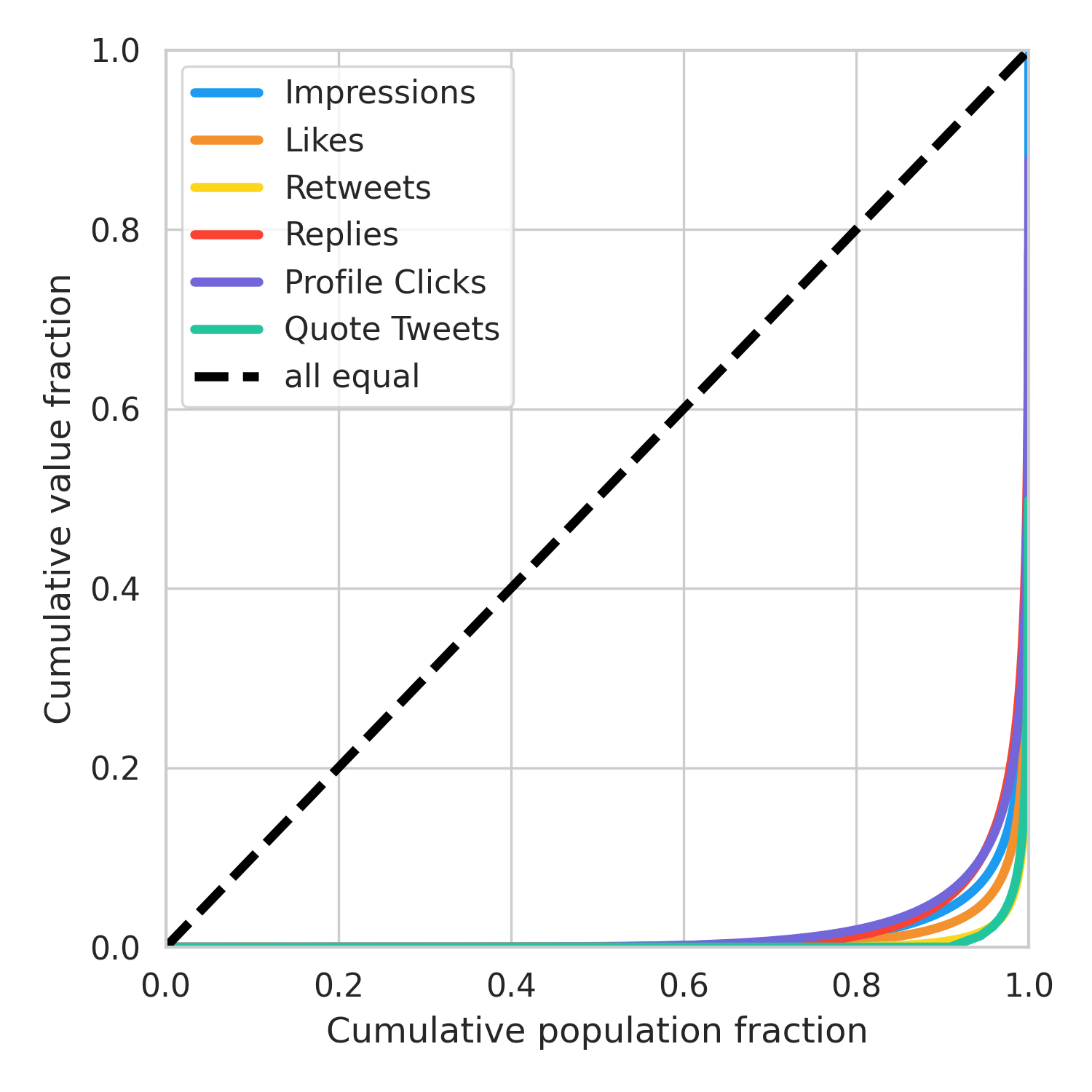}
         \caption{Linear $y$-axis}
         \label{fig:lin_lorenz}
     \end{subfigure}
     \begin{subfigure}[b]{0.4\textwidth}
         \centering
         \includegraphics[width=\textwidth]{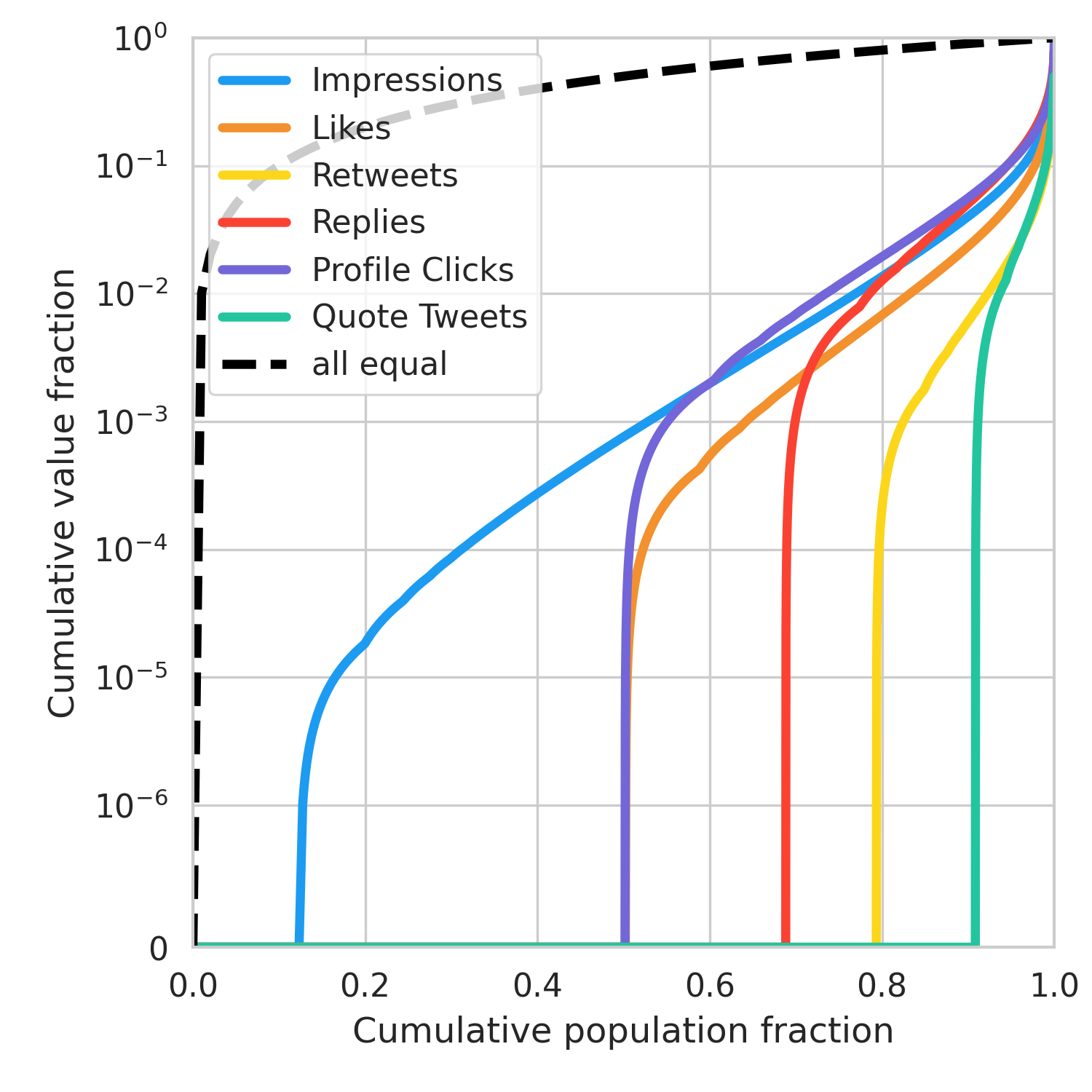}
         \caption{Logarithmic $y$-axis}
         \label{fig:log_lorenz}
     \end{subfigure}
    \caption{Lorenz curves for different types of engagement on Twitter. Lines closer to the dashed black line indicate more equal distributions of engagement. In part~\ref{fig:lin_lorenz}, the more typical linear scale is shown. However, because the distributions are difficult to distinguish, a logarithmic scale is used
  for the $y$-axis in~\ref{fig:log_lorenz}. A small linear portion is included from $0$ to $10^{-6}$ in order to visualize the point at which the distribution transitions from zero to non-zero values.}
    \label{fig:lorenz_engagement}
\end{figure*}

\section{Measuring attention inequality on Twitter}

On social media, number of engagements (interactions with content created by an author) shares many properties with income. Mathematically, it is also a non-negative quantity of which there is a limited amount (largely restricted by reader attention spans). More importantly, though, it can be a measure of the material benefits a person is gaining from the platform. As such, quantifying the imbalance in different types of engagements is a good testing ground for stability, effect sizes, and interpretability.

\subsection{Use case 1: measuring skews in engagements}
\label{sec:engagements}

One way to measure the efficacy of distributional inequality metrics is to use them to quantify how skewed different types of engagements on Twitter are. In this section, we break down the distribution of engagements, both passive and active, finding that the engagements where readers share an author's content to their own followers are the most skewed.

\subsubsection{Dataset and methodology}

To analyze the types of engagement, we first collect a dataset of users who authored Tweets in the month of August 2021. For each of their Tweets in that time frame, we measure a number of different interactions, both passive and active. First, we count a passive interaction, also known as a \textit{linger} \textit{impression} (which we will simply refer to as an impression for short). An impression is logged when at least 50\% of the Tweet is visible for at least 500 ms, indicating that the reader spent some amount of time viewing the Tweet. Next, we consider three types of active click-based engagements: profile clicks, likes, and Retweets. \textit{Profile clicks} happen when a reader clicks on the author's profile icon embedded as part of a Tweet (potentially indicating they wanted to learn more about the author). \textit{Likes} are logged when a reader clicks the heart icon on a Tweet, while \textit{Retweets} are logged when a reader clicks the Retweet icon and shares the Tweet to their own followers. The third and final class of interactions is active content-based engagements. \textit{Replies} happen when a reader clicks the reply icon and posts a Tweet in response to an author's Tweet. \textit{Quote Tweets} occur when a reader shares a Tweet with their own followers, by clicking the Retweet icon, but also adds their own content in response to the Tweet before sharing. These interactions span a range of efforts expended by the reader, from simply spending time on a Tweet to actively replying and sharing it. All interactions are logged by Twitter directly during a user's session. 

The final dataset aggregates, for each Tweet author, the number of interactions over all Tweets composed by that author in August 2021\footnote{The timeframe we use to count the number of impressions is also August 2021. We compared our results to using a timeframe for all impressions received until October 2021 (still using Tweets created in August), but found the differences to be negligible, as the average Tweet receives most of its impressions within a few days of creation.}. It includes authors who posted a Tweet in that month but did not receive any interaction, as long as that author had at least one Tweet selected for display on a timeline. For any particular interaction type, if an author did not receive any engagements of that type from readers in August, their count is zero in that distribution. We restrict our analysis to authors who have at least one follower. After aggregation, the dataset consists of over 100M authors. 

\subsubsection{Results}

Figure~\ref{fig:lorenz_engagement} shows the Lorenz curves for different types of interactions. In~\ref{fig:lin_lorenz}, we see that it is difficult to distinguish the curves on a linear scale, and that the area under the curves is quite small. As such, we show the curves on a logarithmic scale in~\ref{fig:log_lorenz}, with a small linear region from $0$ to $10^{-6}$. This view is interesting because we can clearly see the number of users at which the distribution transitions from zeros to non-zero counts. Unsurprisingly, this transition happens earliest for the impression interactions, as these are the lowest effort of interaction with a Tweet. It is also interesting to note that some of the curves do cross, making it difficult to evaluate the skew of the distributions visually. 

\begin{figure*}
     \centering
     \begin{subfigure}[b]{0.24\textwidth}
         \centering
         \includegraphics[width=\textwidth]{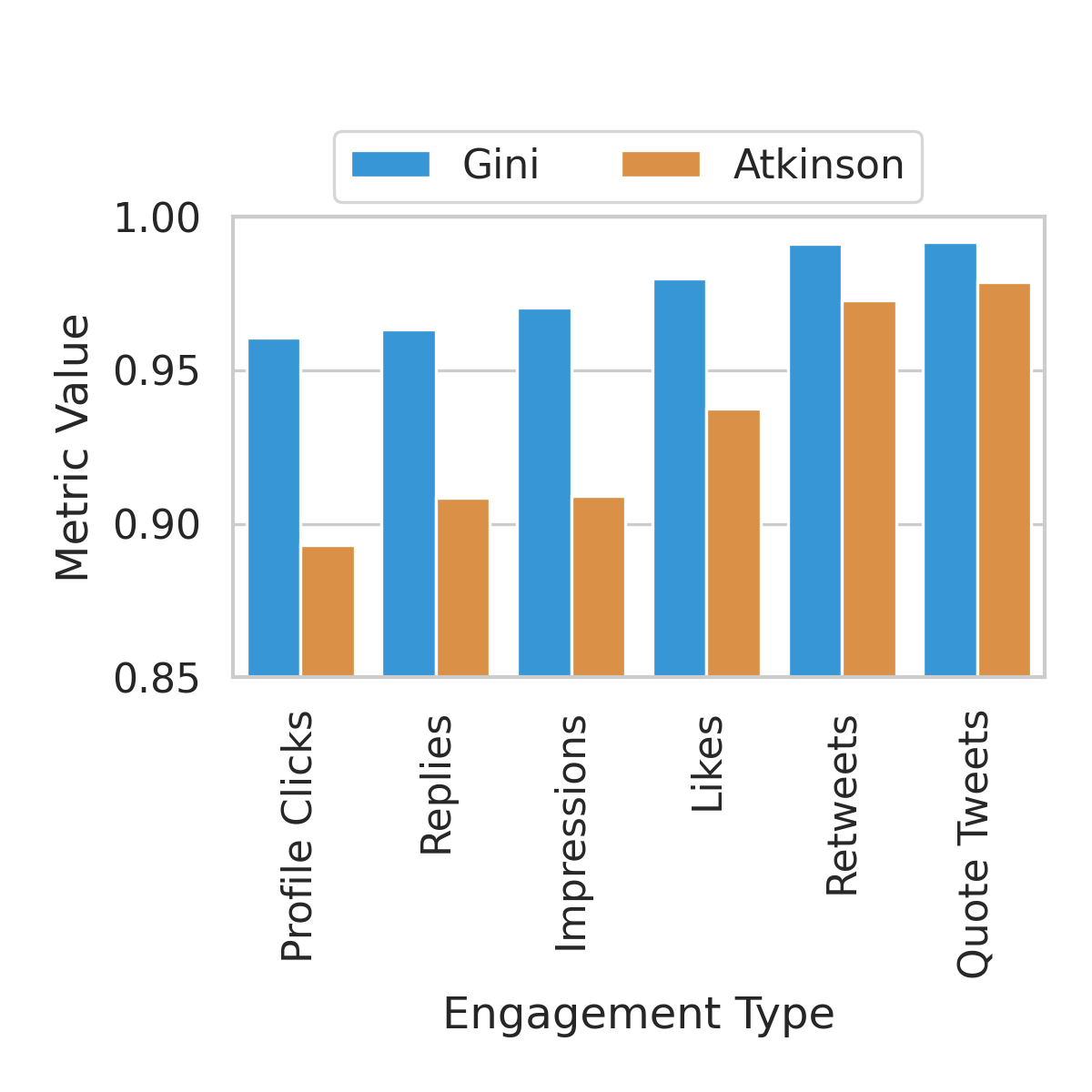}
         \caption{Entropy-based}
         \label{fig:gini_eng}
     \end{subfigure}
     \begin{subfigure}[b]{0.24\textwidth}
         \centering
         \includegraphics[width=\textwidth]{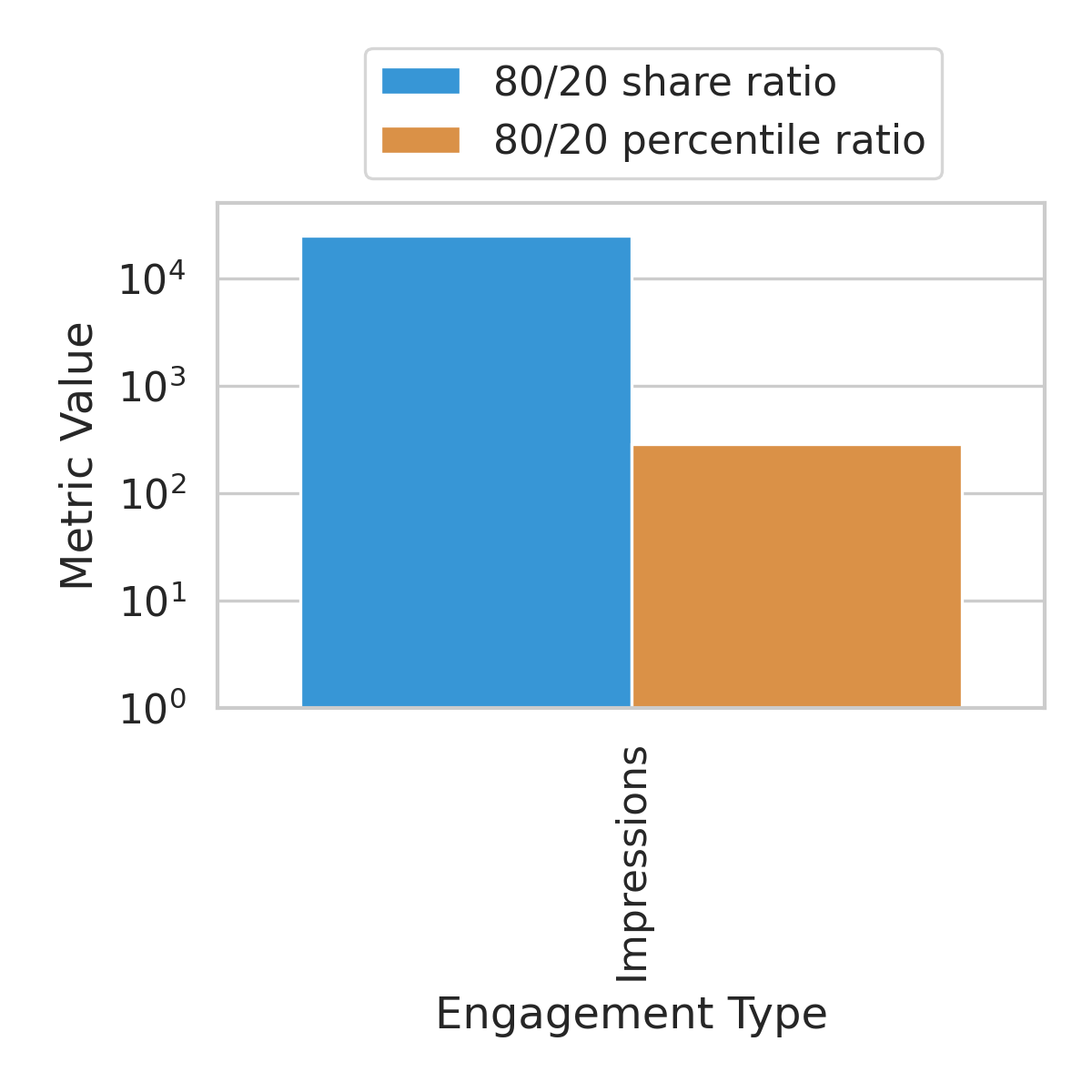}
         \caption{Ratio}
         \label{fig:ratio_eng}
     \end{subfigure} 
     \begin{subfigure}[b]{0.24\textwidth}
         \centering
         \includegraphics[width=\textwidth]{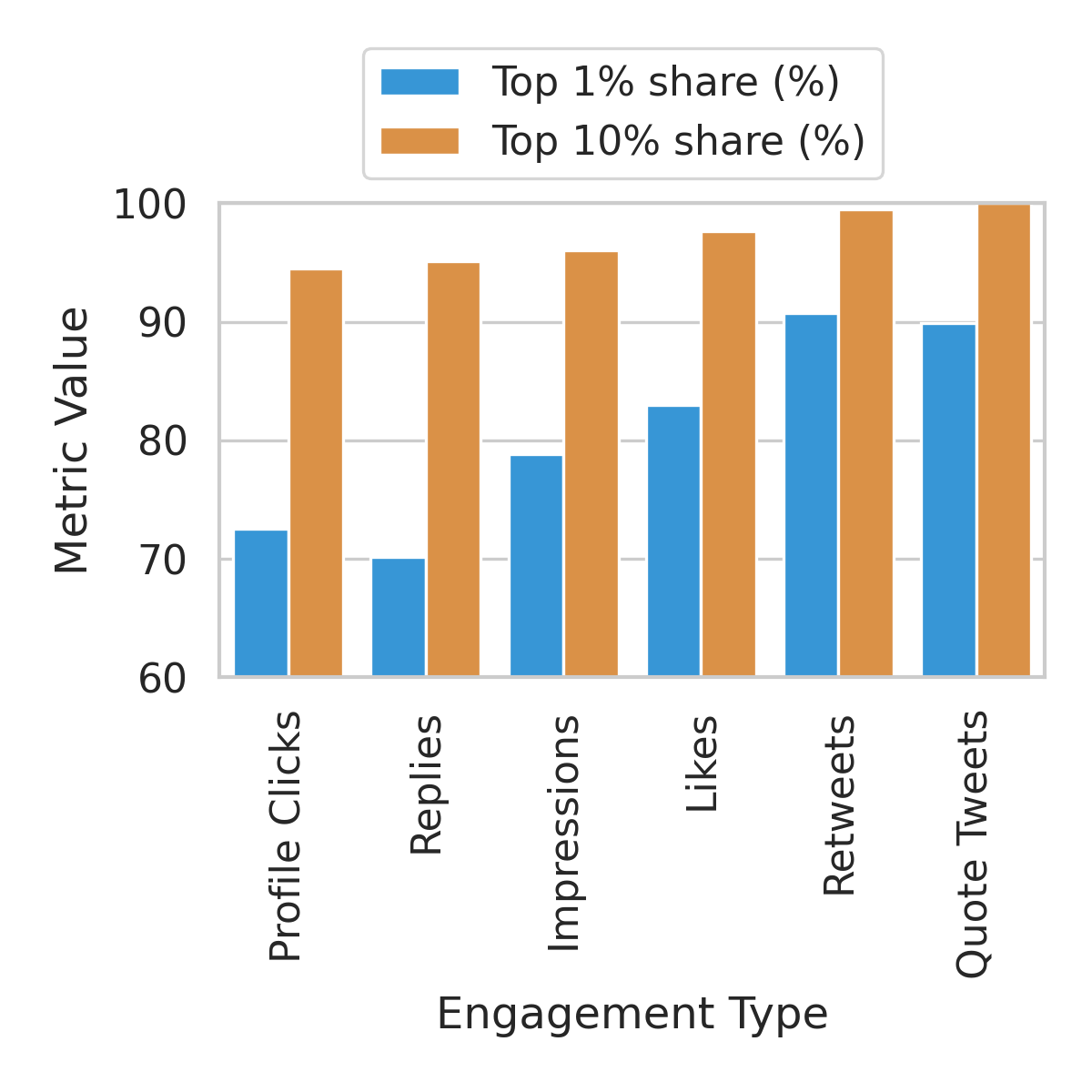}
         \caption{Tail share}
         \label{fig:tailshare_eng}
     \end{subfigure}
     \begin{subfigure}[b]{0.24\textwidth}
         \centering
         \includegraphics[width=\textwidth]{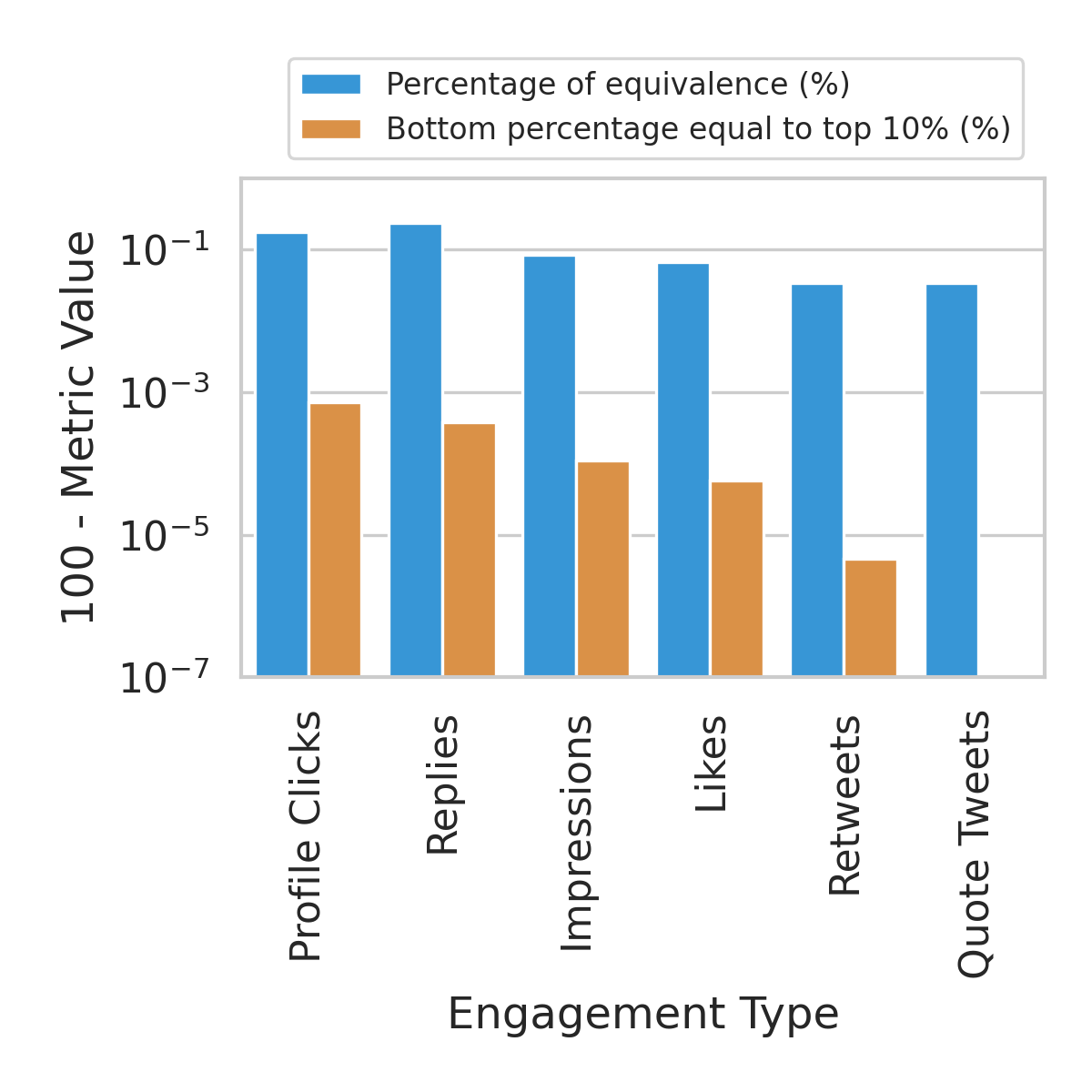}
         \caption{Equivalence}
         \label{fig:equiv_eng}
     \end{subfigure}
    \caption{Families of metrics, computed for the distribution of different engagement types. Figure~\ref{fig:gini_eng} shows the Gini and Atkinson indices. Figure~\ref{fig:ratio_eng} shows the 80/20 share and percentile ratios. Note here that these are only shown for impressions, as for all other types of engagements the share of the bottom 20\% of users is zero. Figure~\ref{fig:tailshare_eng} shows the top 1\% share and top 10\% share for all engagement types. Figure~\ref{fig:equiv_eng} shows the percentage of equivalence and the bottom percentage of users with equal share to the top 10\% of users. Note that in this figure, the percentages are inverted (100 - metric value rather than the value itself) because these metrics are very close to 100\% and are more easily visualized on a logarithmic scale when inverted.}
    \label{fig:metrics_engagment}
\end{figure*}

Figure~\ref{fig:metrics_engagment} shows the breakdown of the distributional inequality metrics by engagement type. For the Atkinson index, we chose $\epsilon=0.5$, as we found this value to give good distinction between the engagement types while still being numerically stable. In each case, the interactions are sorted by their Gini index, to allow for comparison between metrics. To compute errors, we performed bootstrap resampling of the author population and recomputed the metric for each resample, but we found that these error bars were small enough to not be visible (except in the case of figure ~\ref{fig:gini_v_follow}). 

Our first observation is that the metrics all generally agree on the ordering of the distributions, with profile clicks being the least skewed distribution and quote Tweets being the most skewed. We also note that all of the distributions are quite imbalanced, with Gini index above 0.95 in all cases and top 1\% share greater than 70\%. Finally, we note the that 80/20 share and percentile ratios were only calculable for the impression distribution. As can be seen in the Lorenz curves of Figure~\ref{fig:lorenz_engagement}, all other engagements types are still in the zero part of their distribution at the 20th percentile, making the ratio undefined.

In short, the metrics are able to distinguish between imbalances of different types of engagements. Additionally, we find that with very top heavy distributions, ratio metrics are unsuitable as they would need to be adjusted per distribution to find the point where the quantity becomes nonzero in the population. 

\subsection{Use case 2: algorithmic contributors to skew}
\label{sec:suggs}

It is clear from the previous section that engagements with content are quite skewed, with high-ranking authors receiving the bulk of interactions. Tweets can appear on a reader's Home timeline via a variety of different sources. In this section, we propose using the metrics defined above to determine whether certain content suggestion types contribute more strongly to the overall skew of the distribution. This will allow us to determine what effect sizes they have when comparing different algorithm types, as well as giving us a concrete use case in which to evaluate their usability.

\subsubsection{Dataset and methodology} 

For this study, we focus on the impression distribution from the previous section. We break down the number of impressions by the source that recommended that Tweet to the reader, with the source logged by Twitter when a user's timeline is generated during a session. In this dataset, we consider five different categories of suggestions. The first, \textit{In network (IN)} suggestions, consist of Tweets from authors followed by a reader.  Next, we have some aggregated ``out of network" (OON) suggestion types, or suggestions for Tweets that were created by a person not followed by the reader. One type, which we refer to as \textit{OON, Likes}, consists of Tweets that appear on a reader's timeline because they were liked by a person the reader follows, but the reader does not follow the author of the Tweet. Another, \textit{OON, Graph} includes Tweets that are considered interesting based on shared interests among the users, specifically computed using information from the social graph, and include recommendations coming from the SimCluster algorithm~\cite{satuluri2020simclusters}. The final category is an aggregation of miscellaneous other out of network suggestions, including situations where an out of network author was recently followed by in network users, Tweets the reader may be interested based on recent search queries, out of network replies to in network Tweets, and out of network Tweets based on previous entities that the user engaged with. These are collectively referred to as \textit{OON, Misc.}. Finally, we consider a suggestion type which is a hybrid of IN and OON. \textit{Topics} suggestions come from topics that a reader has followed, including cases where the Tweet author themselves is not followed by the reader (though some Tweet authors in the Topic may be followed by the reader). As before, if an author did not receive any impressions via a particular suggestion type, their count is zero for that distribution. 

\subsubsection{Results}

\begin{figure*}
     \centering
     
     \begin{subfigure}[b]{0.3\textwidth}
         \centering
         \includegraphics[width=\textwidth]{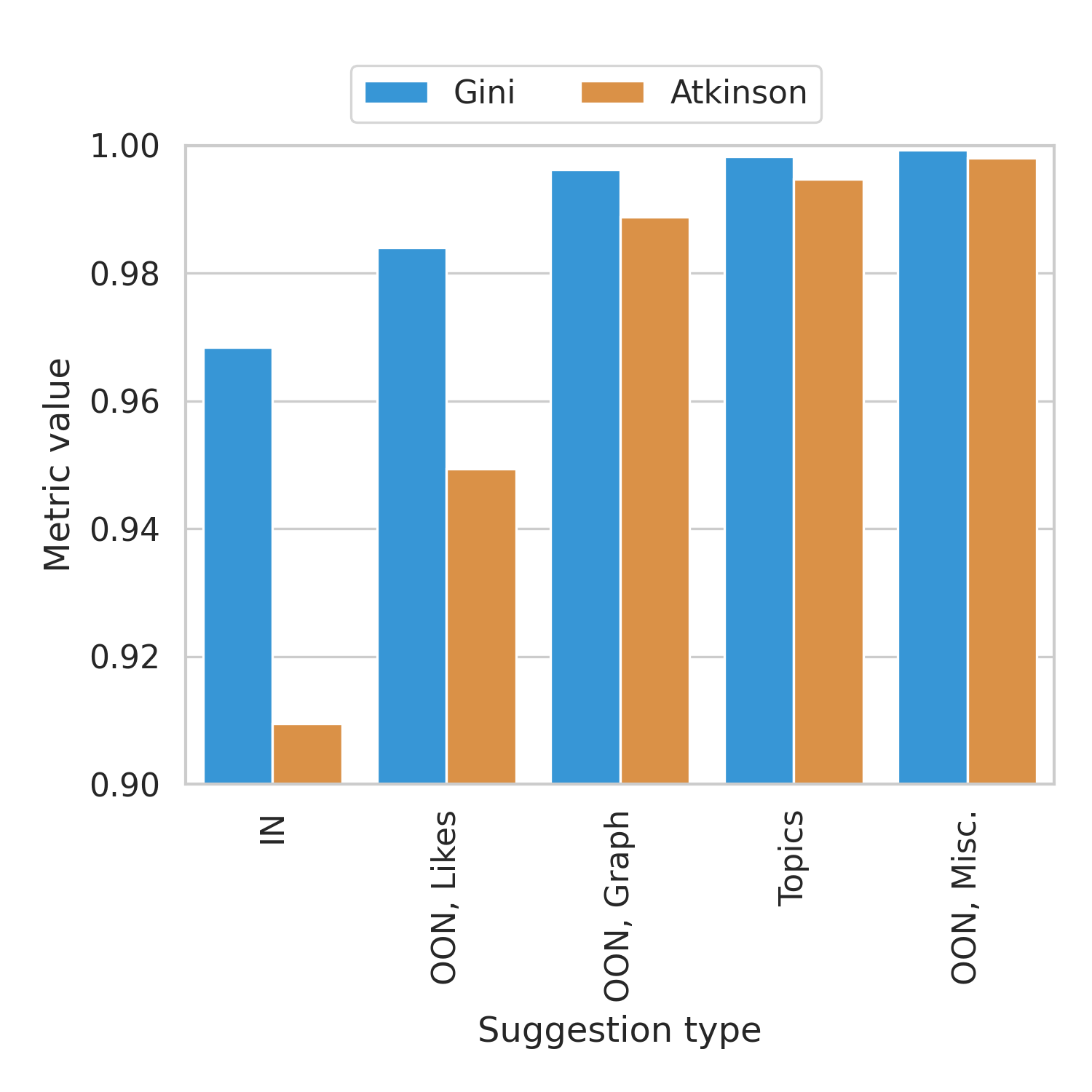}
         \caption{Entropy-based}
         \label{fig:gini_sugg}
     \end{subfigure}
     \begin{subfigure}[b]{0.3\textwidth}
         \centering
         \includegraphics[width=\textwidth]{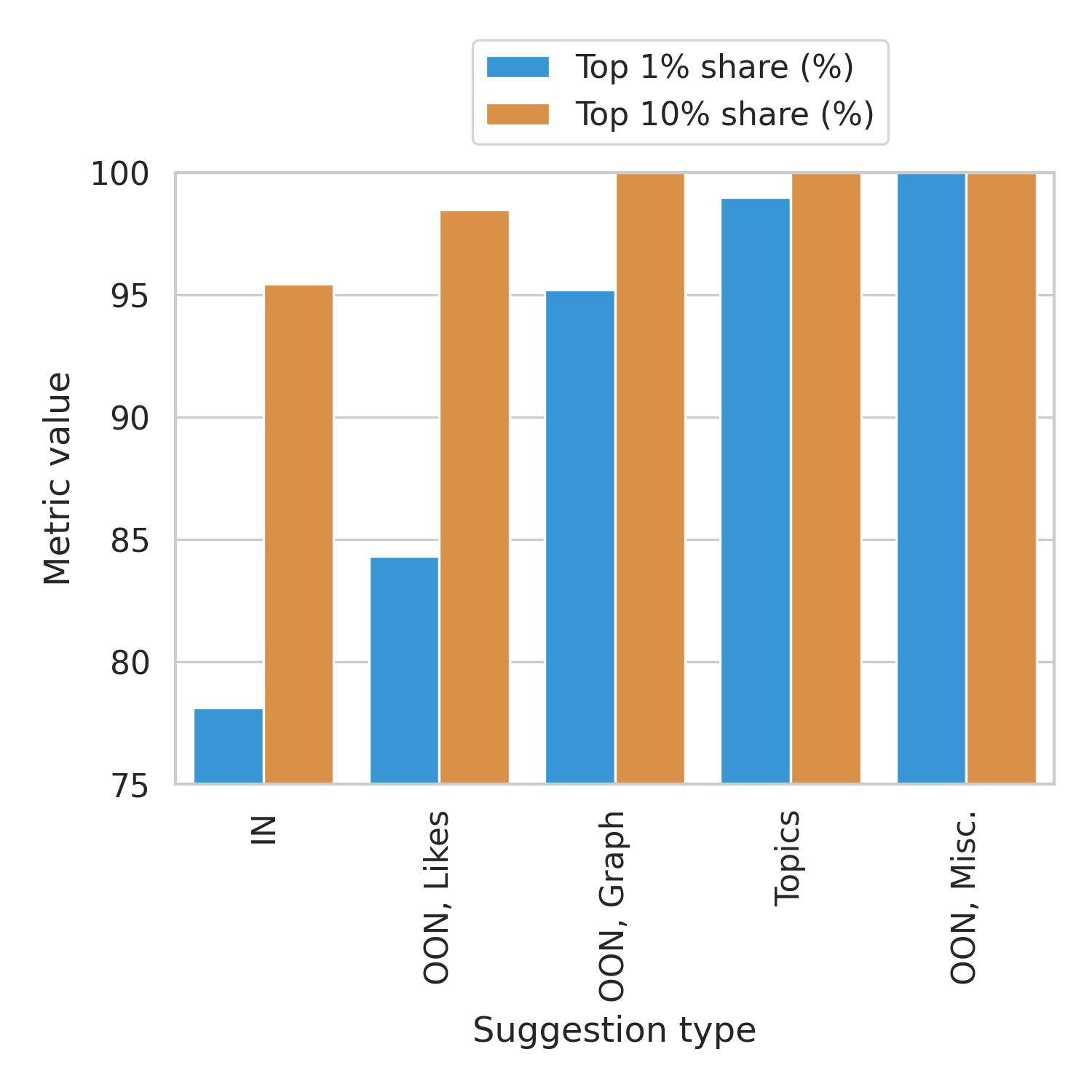}
         \caption{Tail share}
         \label{fig:tailshare_sugg}
     \end{subfigure}
    \caption{Measured values of entropy-based and tail share metrics by suggestion type for the distribution of impressions. We again choose $\epsilon=0.5$ for the Atkinson index, as in figure~\ref{fig:metrics_engagment}. The distribution of impressions from the in network ranking algorithm are the least skewed, while miscellaneous out of network suggestions are the most skewed.}
    \label{fig:metrics_sugg}
\end{figure*}

Figure~\ref{fig:metrics_sugg} shows the breakdown of metrics for  impressions by suggestion type. Here, we only report the entropy and tail share metrics, as the others are ill-defined due to the presence of many zeros in the distributions. We find generally that suggestions for in network Tweets have more equitable distributions of impressions than out of network Tweets. This is likely due to the fact that the OON suggestion algorithms rely on the structure of the social graph. For example, in the case of \textit{OON, Favorites}, the author must have a path to a reader B  through follower A in order to appear on reader B's feed. For in network suggestion types, the number of impressions should grow roughly linearly with the number of followers, as these suggestions only apply to readers following the author. However, OON suggestion types are likely to have their impressions grow faster than linearly, as adding a single follower also adds paths to that follower's followers. 

\begin{figure}
     \centering
     \includegraphics[height=0.3\textheight]{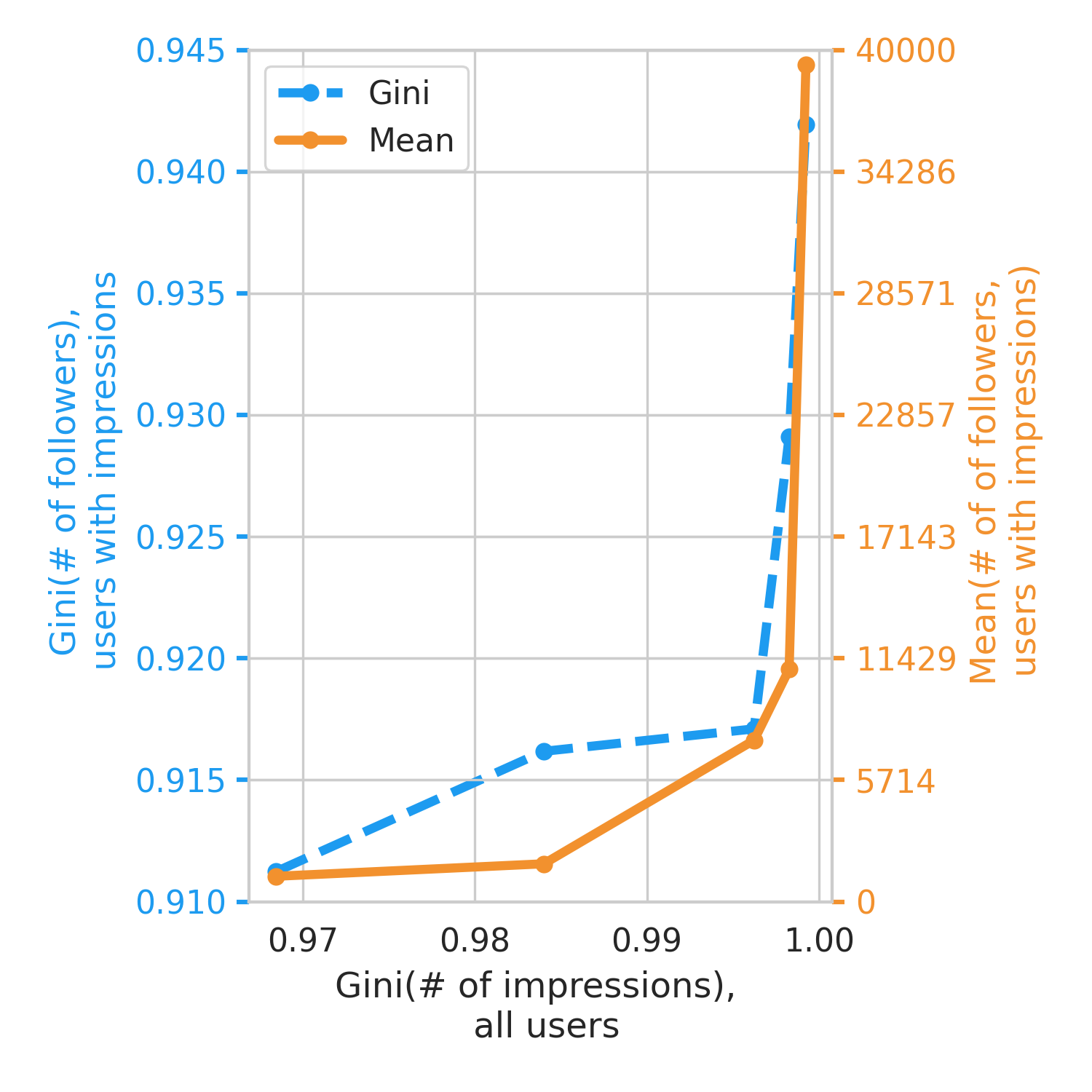}
     \caption{A comparison of Gini coefficient of number of impressions and statistics of number of followers. Each point is one suggestion type, and the $y$-axis shows the Gini index and average of number of followers for users who received impressions from that suggestion type.}
     \label{fig:gini_followers}
\end{figure}

To further analyze these results, we consider the relationship between impression skew and number of followers, focusing on the Gini index\footnote{Gini, Atkinson, and top 1\% share were all found to be highly correlated in this data, so we chose Gini index for simplicity}. Figure~\ref{fig:gini_followers} shows the Gini index and average of the number of followers distribution for authors who received impressions from a particular suggestion type as a function of the Gini index for that suggestion type from all users (the quantity shown in figure~\ref{fig:gini_sugg}). We see that suggestion types which have larger Gini indices for impressions also have larger means and Gini indices for  number of followers, indicating that authors who benefit from the more skewed interaction types have more followers. 

\begin{figure*}
     \centering
     
     \begin{subfigure}[b]{0.4\textwidth}
         \centering
         \includegraphics[width=\textwidth]{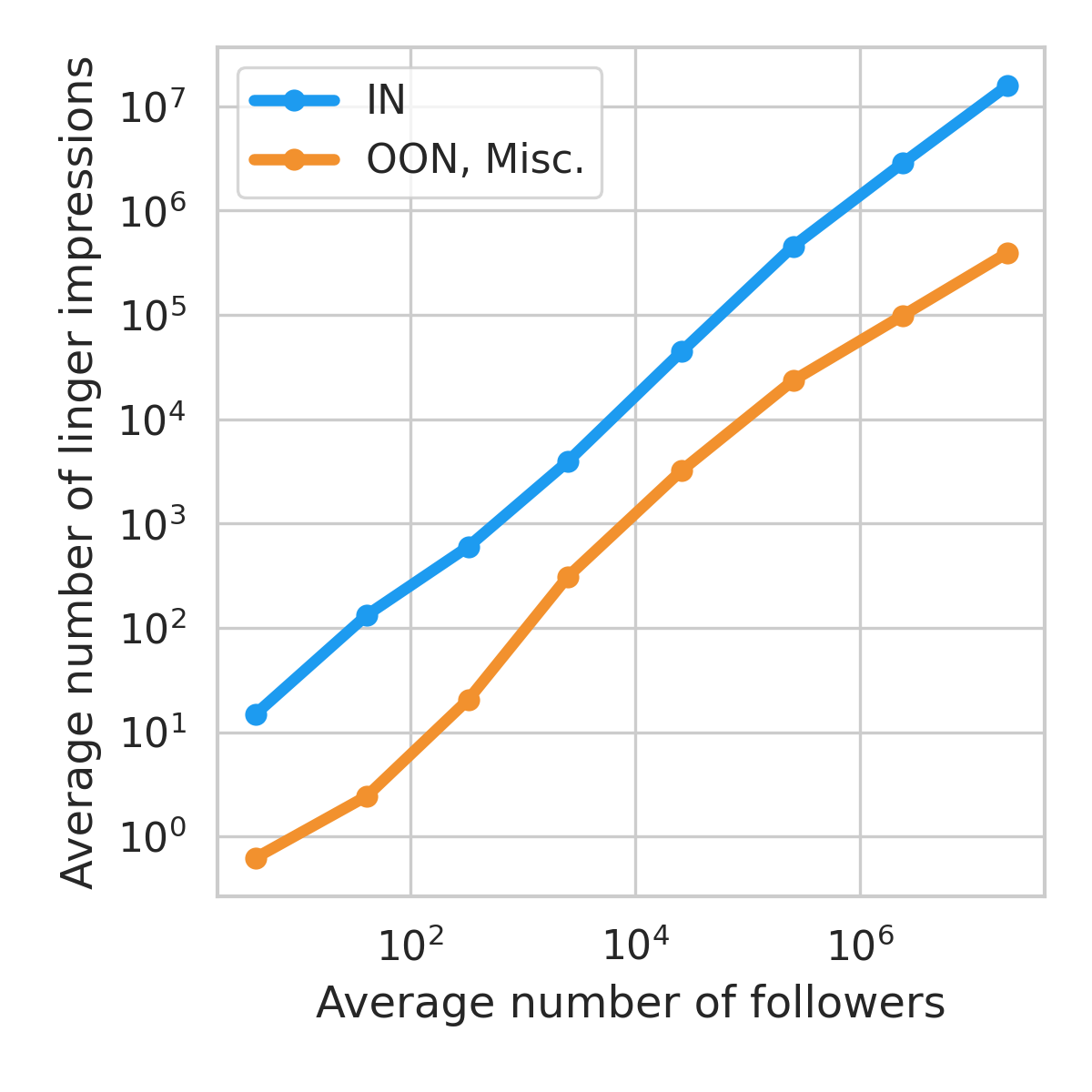}
         \caption{Average impressions vs. followers}
         \label{fig:impress_v_follow}
     \end{subfigure}
     \begin{subfigure}[b]{0.4\textwidth}
         \centering
         \includegraphics[width=\textwidth]{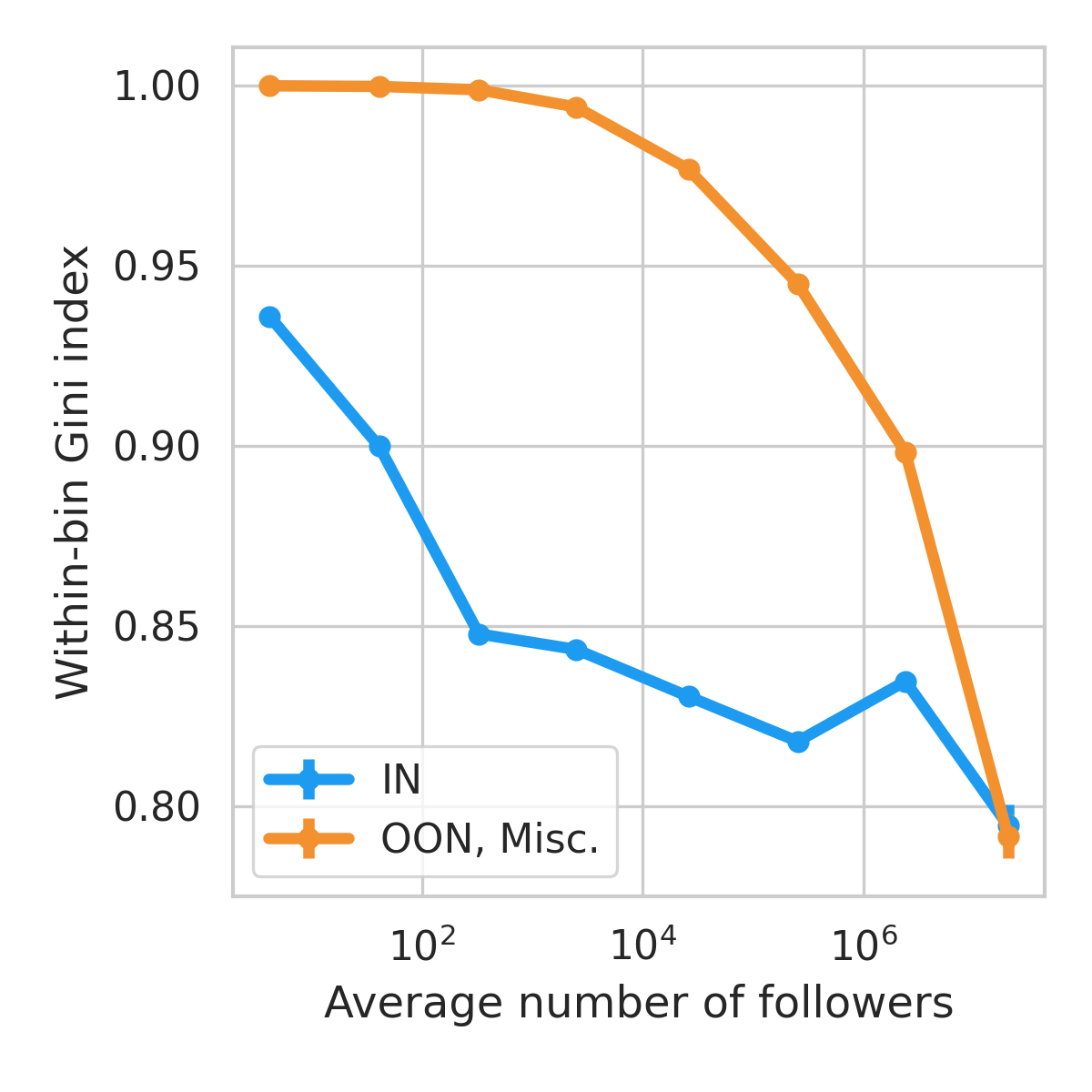}
         \caption{Gini vs. followers}
         \label{fig:gini_v_follow}
     \end{subfigure}
     
    \caption{Breakdown of the ranking and miscellaneous out of network suggestion types by number of followers. Figure~\ref{fig:impress_v_follow} shows that the average number of impressions increases similarly with number of followers for both algorithms, with the overall number of ranking impressions being larger. Figure~\ref{fig:gini_v_follow} shows that, for authors with lower numbers of followers, the distribution of impressions from out of network sources is significantly more skewed than for in network sources. In the highest bin, the Gini indices between the two sources are close to another, showing that the the distributions of impressions are very similar for both suggestion types once an author has enough followers.}
    \label{fig:money_plot}
\end{figure*}

Another way to see this effect is shown in figure~\ref{fig:money_plot}. Here, we bin authors by number of followers, allowing us to compute the within-bin Gini indices and average number of impressions by bin. Figure~\ref{fig:impress_v_follow} shows how the average number of impressions grows with average number of followers in the bin. The number of impressions grows at a similar rate for both for the ranking and OON suggestions, with authors in general getting more impressions via the in-network ranking suggestions. However, when we look at the within-bin Gini index versus the average number of followers, we see that the distribution of impressions that come from OON suggestions is much more skewed for authors with low number of followers. The within-bin Gini indices for authors with very high number of followers are almost identical between the ranking and OON suggestions, while for lower numbers of followers the OON distribution is significantly more skewed. This gives some support to the hypothesis that the larger disparities seen in OON suggestion types are driven by the number of followers an author has, with a smaller fraction of low-follower authors getting an opportunity to have their content exposed via OON suggestions. We note also that some of these disparities may be coming from biases in upstream algorithm behavior as well, and an interesting direction for future work would be to attempt to decompose the influence of the algorithm itself from the structure of the network.

Overall, these results demonstrate the richness of insights that can be derived from the use of these metrics in operational settings. Additionally, they have given us a better understanding of how they satisfy the qualitative and empirical properties we set forth, as we discuss in more detail in the next section. 

\section{Discussion: How do the proposed metrics measure up?}
\label{sec:evaluation}

In the empirical use cases above, we have seen that distributional inequality metrics can help us glean detailed information about both the state and origins of skews algorithm-driven outcomes on a platform like Twitter. In this section, we will revisit the proposed criteria we set forth early on in light of those analyses. 

\subsection{Evaluation with respect to desirable criteria}

\begin{table*}[h]
\centering
\begin{tabular}{|c||c|c|c|c|c|c|c|c|}
\hline
& \multicolumn{4}{|c|}{\bf Theoretical} & \multicolumn{2}{|c|}{\bf Qualitative} & \multicolumn{2}{|c|}{\bf Empirical} \\ \hline
& Population? & Adjustable? & Scale? & Subgroups? & User? & Interpretable? & Stable? & Effect? \\ \hline
Gini & \yes & \no & \yes & \yes & \no & \low & \high & \medium \\\hline
Atkinson & \yes & \yes & \yes & \yes & \no & \low & \high & \medium \\\hline
\textbf{Top X\% share} & \textbf{\yes} & \textbf{\yes} & \textbf{\yes} & \no & \yes & \textbf{\high} & \textbf{\high} & \textbf{\high} \\ \hline
Percentile ratio & \yes & \yes & \yes & \no & \yes & \medium & \low & \medium \\ \hline
Share ratio & \yes & \yes & \yes & \no & \yes & \medium & \low & \medium \\ \hline
Equiv. to top X\% & \no & \yes & \yes & \no & \yes & \medium & \high & \low \\ \hline
\% of equal share & \no & \no & \yes & \no & \yes & \high & \high & \low \\ \hline
\end{tabular}
\caption{Summary of metrics and desirable criteria}
\label{tab:metric_evaluation}
\end{table*}

Table~\ref{tab:metric_evaluation} shows our evaluation of the criteria based on the analysis of impression distributions we conducted. 

With respect to the theoretical criteria, the performance of the metrics varies. The entropy, tail share, and ratio metrics are all population invariant, meaning that they have the same meaning regardless of population size. The equivalence metrics do not satisfy this criterion, as the percentage they output has a different meaning in terms of number of users as the population changes. For adjustability, most metrics had an adjustable parameter, with the exception of Gini index and percentage of equal share. All of the metrics were multiplicative scale invariant, keeping the same output if all values in the distribution are multiplied by a constant. Finally, only the entropy metrics were subgroup decomposable, with all the other metrics lacking this property because the percentages calculated on subgroups are not easily mapped to the full population due to overlaps in ranks between the groups (e.g. a person in the top 1\% of a subgroup could be in the bottom 1\% of the full population)\footnote{We note that Atkinson specifically is \textit{additively decomposable}, meaning that the total Atkinson index is a sum of the Atkinson index on subgroups. Gini is still decomposable, but the total Gini is a weighted average of subgroup Gini indices, with the weights being the population fraction of the subgroup.}. 

For the qualitative criteria, we found a similar mix of results. In user focus, all percentage-based metrics passed the criteria, as either their parameters or their outputs could be directly translated to a number of users. The entropy metrics did not meet this criterion. For interpretability, the most interpretable metrics were the top X\% share and the percentage of equal share. We rate these metrics as highly interpretable because they can be phrased in ways used in everyday discussion of income, e.g. ``the top 1\% of authors get 95\% of all impressions'' or ``the top 30\% of authors have the same share as the bottom 70\%". The ratio metrics have medium interpretability because they do still map to sections of users in the distribution, but it is harder to translate the ratios into statements about the individual segments. Finally, we rate the entropy metrics as having low interpretability, since a difference in Gini or Atkinson of some amount does not have an intuitive framing for practitioners in terms of numbers of people or redistribution of value. 

Finally, we evaluate the empirical criteria based on the use cases examined earlier in the paper. All metrics except the ratio metrics are highly stable, with negligible error bars seen from bootstrap resampling. We rate the ratio metrics as having low stability because their values can become quite large if the low end of the population does not have a large share of wealth, and in many cases they can even be undefined. For effect size, the top X\% share had the highest rating, with a large dynamic range when comparing different engagement types or algorithms (see figures~\ref{fig:tailshare_eng} and~\ref{fig:tailshare_sugg}). The entropy and ratio metrics did not show as much spread between different distributions as the tail share and thus were rated as medium. The equivalence metrics received low effect size ratings, as these metrics had to be visualized inversely on a log scale in order to distinguish differences between the distributions. 

When considering all the criteria together, we find that the top X\% share received the highest ratings for this dataset, only failing the subgroup decomposability criterion. The entropy-based metrics had low interpretability and no user focus, but did well in other metrics. Ratio and equivalence metrics had low stability and effect size, respectively, and failed some of the theoretical criteria. Overall, we hope that these evaluations will be useful for practitioners in deciding which metric may be best suited for their use case. 

\subsection{Limitations of proposed metrics}

While we found that distributional inequality metrics were useful for capturing discrepancies in exposure for authors on Twitter, these metrics (like any metric) do have inherent limitations. First, these metrics are most useful for distributions where ``improvement" can be defined as a reduction in the skew. If, for example, these metrics were applied to a quantity where benefit vs. harm is more ambiguous, they might be less useful in evaluating disparities. Second, while their lack of reliance on demographic information is useful in an operational context, they cannot ever fully replace measurements of disparate treatment or outcomes between groups. They are meant as a supplement to illustrate tradeoffs rather than a replacement, as it will always be crucial to understand how algorithms are affecting underrepresented groups specifically. Finally, our evaluation of the criteria is limited to the characteristics of the datasets we evaluated. However, we feel that these highly concentrated impressions distributions are indicative of many distributions on internet datasets, and therefore these metrics should be useful for any distributions that exhibit similar properties. 

\subsection{Future directions}

Based on what we learned from the evaluation of these metrics, we have found a few key paths for future work that we intend to pursue. First, we are developing tools to leverage inequality metrics during product development, so that those building ML models can use them to understand disparate impacts. Second, we aim specifically to understand how to integrate these metrics into A/B  testing  frameworks, particularly focusing on how the choice of  randomization unit affects the choice of metric. Finally, we believe more work is needed to understand the changes in inequality as a function of number of followers for in-network vs. out-of-network suggestions. To this end, future work could focus on measures of inequality that explicitly incorporate information about graph structure. 

\section{Conclusion}

In this paper, we evaluated a number of metrics in the context of outcomes of Twitter's recommendation system. We found that certain metrics are useful in different contexts, with the top X\% share performing the best according to our evaluation criteria. We use these metrics to identify sources of skew in engagements with content on Twitter, particularly noting that certain out of network suggestions lead to more skewed outcomes. In addition, we show that having a lower number of followers disproportionately skews outcomes for out of network suggestions compared to in network suggestions. Overall, we find that these metrics are useful tools for identifying algorithmic sources of disparate outcomes.


\bibliography{main}

\appendix

\section{Supplementary material}

\begin{figure}
  \centering
  \includegraphics[width=0.9\linewidth]{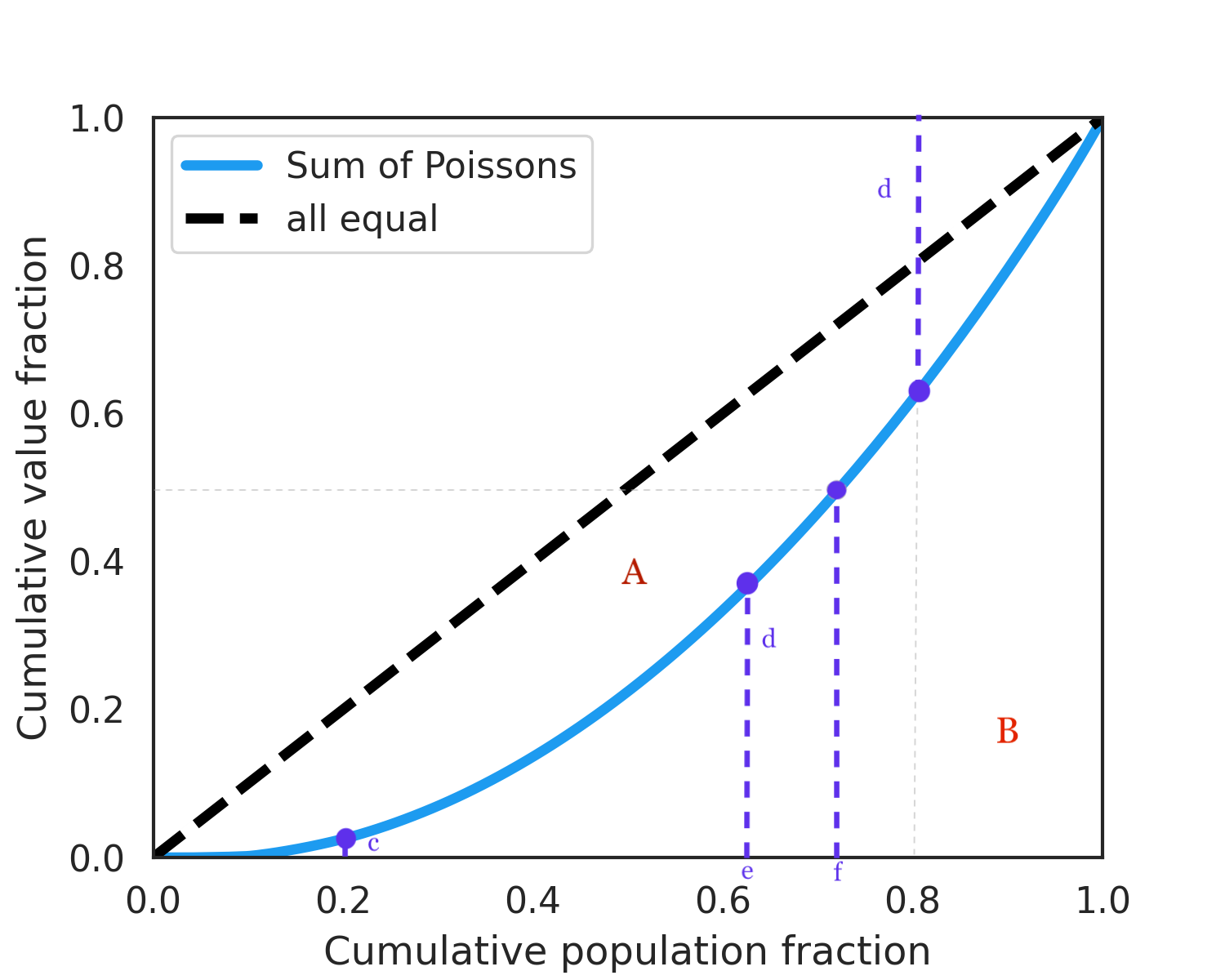}
  \caption{An example Lorenz curve with annotations that can be used to derive several related metrics. Capital letters ($A$ and $B$) are areas, while the rest are lengths}
  \label{fig:annotated_lorenz}
\end{figure}

\subsection{Visual representation of inequality metrics}
\label{sec:visual_lorenz}

Figure~\ref{fig:annotated_lorenz} shows an annotated Lorenz curve of synthetic data (generated from a sum of Poisson distributions) that we can use to define other quantities of interest. The follow metrics can be defined in terms of quantities labeled on the curve:

\begin{itemize}
    \item Gini index: $A$ is the area between the curve and the line of equality, and $B$ is the area below the Lorenz curve. The Gini index can be defined as $A/(A+B)$. Equivalently, because $A+B=0.5$, it is equal to $2A$ or $1-2B$. 
    \item Top X\% share: $d$ is 1 minus the value of the curve at a fraction of 0.8. This corresponds to the share of the top 20\% of individuals in the population.
    \item Share ratio: $c$ is the value of the Lorenz curve at a cumulative population fraction of 0.2. The quantity $d/c$ is the 80/20 share ratio, or the share held by the top 20\% divided by the share held by the bottom 20\%.
    \item Equivalent to top X\%: The cumulative population fraction $e$ is the bottom percentage of users equivalent to the top 20\%, as it is the fraction at which the point on the Lorenz curve equals $d$.
    \item Percentage of equivalence: The fraction $f$ is the point at which the Lorenz curve's value is 50\%, meaning that the top $100f$\% of individuals has the same share as the bottom $100(1-f)$\%. 
\end{itemize}

\subsection{Distribution of followers}

\begin{figure}
    \centering
    \includegraphics[width=0.7\linewidth]{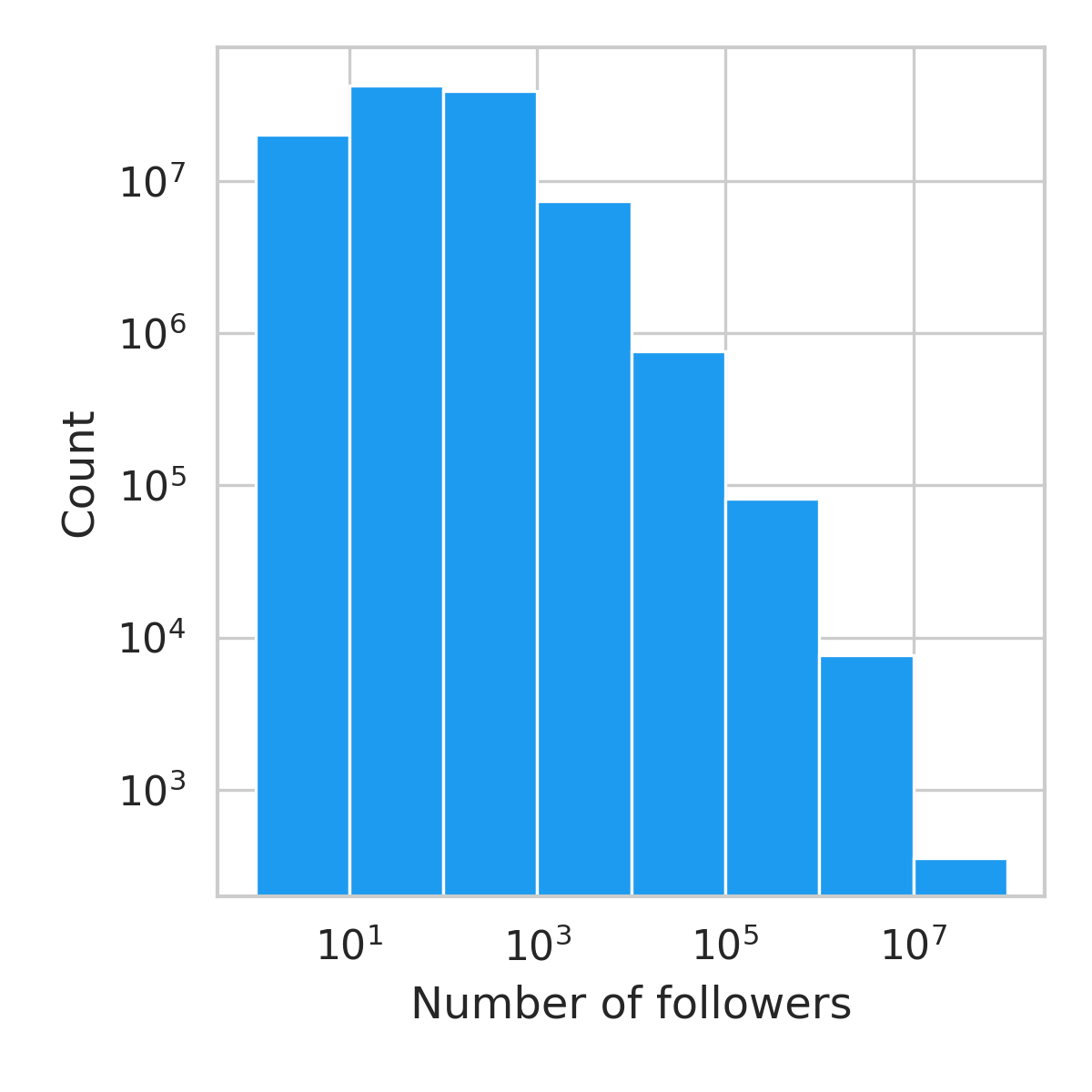}
    \caption{Distribution of number of followers for users in the dataset described in the results sections. The bins here correspond to the bins used to define each point in figure~\ref{fig:money_plot}.}
    \label{fig:follower_dist}
\end{figure}

In the second case study, we explored the impressions distributions and metrics as a function of number of followers. Figure~\ref{fig:follower_dist} shows the distribution of number of followers, as was used for binning in that section.

\end{document}